\documentclass{article}

\usepackage{PRIMEarxiv}

\usepackage[utf8]{inputenc} 
\usepackage[T1]{fontenc}    
\usepackage{hyperref}       
\usepackage{url}            
\usepackage{booktabs}       
\usepackage{amsfonts}       
\usepackage{nicefrac}       
\usepackage{microtype}      
\usepackage{lipsum}
\usepackage{fancyhdr}       
\usepackage{graphicx}       
\graphicspath{{media/}}     
\usepackage[T1]{fontenc}
\usepackage{graphicx}
\usepackage{authblk}
\usepackage{blindtext}

\title{3D-Morphomics, Morphological Features on CT scans for lung nodule malignancy diagnosis}
\author[1, 2]{Elias Munoz}
\author[1, 3]{Pierre Baudot}
\author[1]{Van-Khoa Le}
\author[1]{Charles Voyton }
\author[1]{Benjamin Renoust}
\author[1]{Danny Francis}
\author[1]{Vladimir Groza}
\author[1]{Jean-Christophe Brisset}
\author[1]{Ezequiel Geremia}
\author[1]{Antoine Iannessi}
\author[1]{Yan Liu}
\author[1]{Benoit Huet}
\affil[1]{Median Technologies, 1800 Rte des Crêtes Batiment B, 06560 Valbonne, France.}
\affil[2]{Univ. Artois, UR 2462, Laboratoire de Mathématiques de Lens (LML) F-62300}
\affil[3]{Correspondence: pierre.baudot@mediantechnologies.com}

\begin{document}
\maketitle

\begin{abstract}
Pathologies systematically induce morphological changes, thus providing a major but yet insufficiently quantified source of observables for diagnosis. The study develops a predictive model of the pathological states based on morphological features (3D-morphomics) on Computed Tomography (CT) volumes. A complete workflow for mesh extraction and simplification of an organ's surface is developed, and coupled with an automatic extraction of morphological features given by the distribution of mean curvature and mesh energy. An XGBoost supervised classifier is then trained and tested on the 3D-morphomics to predict the pathological states. This framework is applied to the prediction of the malignancy of lung's nodules. On a subset of NLST database with malignancy confirmed biopsy, using 3D-morphomics only, the classification model of lung nodules into malignant vs. benign achieves 0.964 of AUC. Three other sets of classical features are trained and tested, (1) clinical relevant features gives an AUC of 0.58, (2) 111 radiomics gives an AUC of 0.976, (3) radiologist ground truth (GT) containing the nodule size, attenuation and spiculation qualitative annotations gives an AUC of 0.979. We also test the Brock model and obtain an AUC of 0.826. Combining 3D-morphomics and radiomics features achieves state-of-the-art results with an AUC of 0.978 where the 3D-morphomics have some of the highest predictive powers. As a validation on a public independent cohort, models are applied to the LIDC dataset, the 3D-morphomics achieves an AUC of 0.906  and the 3D-morphomics+radiomics achieves an AUC of 0.958, which ranks second in the challenge among deep models. It establishes the curvature distributions as efficient features for predicting lung nodule malignancy and a new method that can be applied directly to arbitrary computer aided diagnosis task. 
\end{abstract}

\keywords{mesh  \and radiomics \and Computed Tomography \and Lung cancer screening \and Computer Aided Diagnosis \and Computational anatomy.}

\section{Introduction}

 Since the beginning of medicine, morphological characteristics have provided phenotypes of symptoms allowing to establish the clinical diagnosis. According to the generic principle of a correspondence between structure and function in biology, any dysfunction of a biological process goes in hand with a pathological deformation of the underlying biological structure. Unfortunately, morphological features are often difficult to quantify; at best limited to coarse quantifiers, if not purely qualitative observations. As a consequence, most of the observations of morphological features are not reliable in clinical practice, and are progressively replaced by biomarkers. The aim of this study is twofold: first, the development of a basic and generic method of morphological classification based on the extraction of mesh curvature distribution; second, a validation of the method on clinical prediction of lung nodules malignancy. Curvature is one of the main topological invariant, in the sense of the Bonnet-Gauss theorem, and hence a robust descriptor of the shape of manifold. However, in order to obtain feature that are sensible to deformations and hence not diffeomorphic invariant, we consider the distribution of curvatures which encodes a wide spectrum of deformations. For a theoretical context, we refer to Federer \cite{federer_curvature_1959} and note that we consider the Bonnet-Gauss theorem as giving a signed measure, which distribution of its atomic element (at each vertex) defines our set of features.\\
Non-invasive automatic diagnosis of disease on CT images are based on different kinds of models illustrated here in the context of lung cancer: 
\begin{itemize}
    \item (1) size features-based models (basic morphology) as in the popular LungRADS model \cite{mckee_performance_2015},
    \item (2) clinical features-based model like PANCAN \cite{tammemagi_participant_2017} that mostly relies on patients' clinical and historical information like age, antecedent and smoking habits sometime associated with size \cite{huang_prediction_2019},
    \item (3) radiomics features-based models, commonly textural and image statistics features \cite{ranjbar_chapter_2017,aerts_decoding_2014,wilson_radiomics_2017},
    \item (4) deep network models such as convolutions nets \cite{ardila_end--end_2019} or attention-based models \cite{al-shabi_procan_2022}, that provide the best performances and are theoretically optimal. 
\end{itemize}

Radiomics most often rely on textural features, luminance intensity statistics features, and in some case to basic shape quantification such as 2D ellipse's radius. In the context of lung cancer, the clinician proposes his diagnosis of nodule mostly based on the size of the nodule, along with two sets of qualitative features representing the shape of the nodule (margins, contours) and the luminance profile of the nodule (calcification, attenuation) \cite{erasmus_solitary_2000}. Notably, spiculations and lobulations, which are spikes or bumps on the surface of the pulmonary nodules, as opposed to the "sphericity", are important predictors of the lung nodule malignancy.\\
In an unsupervised sub-classification study of lung or head-and-neck tumors, Aerts et al. \cite{aerts_decoding_2014} designed a state-of the-art library of radiomics including image intensity, texture, but also introduced shape features like compactness or sphericity computed as functions of surface to volume ratios \cite{van_griethuysen_computational_2017}, along with the more classical size and volume features. Their sets of 110 features will be used as a baseline of the present study. \\
More refined morphological study is realized by Leonardi et al.\cite{leonardi_multiple_2015}, that proposes to localize kidney exophytic tumors on a surface mesh using a maximum curvature or a recursive labelling of the vertices method. Also remarkably, Choi et al. quantify spiculation by reconstructing a 3D mesh from a mask of a nodule, and conformally map it to a sphere and then measure the area distortion \cite{choi_reproducible_2021}. Spiculations may be identified by the set of minima of area distortion. Completing this spiculation feature with size, attachment, and lobulation features, the authors obtained an AUC of 0.82 with a training and test on LIDC dataset.

\section{Methods}
\subsection{Data sets}
\subsubsection{NLST dataset} 
The National Lung Screening Trial (NLST) is a US cancer screening program with 7 years follow-up study with yearly survey \cite{jemal_lung_2017,noauthor_reduced_2011}, with at most the first 3 years CT, available publicly on request. We only consider the 618 cancer patients, and 1201 non-cancer patients with nodules taken randomly among the 8210 with 3 time points eligible for download, with Low Dose CT scans.For each time points, a single CT scan is selected among the multiples CT scans kernels available using the same criterion as Ardila et al. (\cite{ardila_end--end_2019}. \\

\begin{table}
\centering
\caption{Numbers of lesions for each set and subsets of the study.
tables.}\label{tab1}
\begin{tabular}{|l|l|l|l|}
\hline
Dataset &  \# benign nodule & \# malignant nodule & total \# nodule\\
\hline
NLST train-validation set  & 11304 & 372 & 11676\\
NLST test set & 4422 & 151 & 4573\\
NLST total & 15726 & 523 & 16249\\
LIDC independent test & 304 & 352 & 656\\
\hline
\end{tabular}
\end{table}

The annotations, conducted by two expert radiologists, consisted in two tasks: (1) a semi-manual segmentation of all segmentable nodules and (2) a disambiguation task that associate each segmented nodule with its NLST GT that contains notably the nodules result of biopsy and localization, and to each new detected lesion not in NLST GT with a new GT. 523 cancer patients have at least one identified malignant lesion identified by biopsy. Only solid and part-solid parenchymal nodules are selected. The set of malignant lesions is defined as the lesion identified by biopsy as malignant at the time of the cancer diagnosis. The set of benign lesions is defined as all the selected lesion at all time points of non-cancer patients, and the set of calcified lesions in cancer patients that are known to be benign nodules. The remarkable aspect of this database relies in the quality of the malignancy GT: the malignancy of lesions is confirmed at the histological level by invasive procedure. The non-cancer status of patients and nodules is confirmed by up to 7 years follow up. This selection hence includes all the False Positive (FP) of clinician diagnosis (biopsied lesions confirmed benign by histology) as well as all False Negative (FN) of clinicians (not biopsied lesions).\\
The data set is split randomly at the patient level into a train+validation set and a test set containing 168 cancer and 330 benign patients. The resulting number of nodules of each set is resumed in Table~\ref{tab1}: the lesion inclusion-exclusion resulted into a total of 523 malignant lesions and 15726 benign lesions.

\subsubsection{LIDC-IDRI independent test cohort} 
The Lung Image Database Consortium (LIDC-IDRI) database is a public and multicentric US lung cancer screening database \cite{armato_lung_2011} of 1018 CT scans from 1010 patients. For each patient, the dataset includes a CT scan, the annotations and segmentations performed by up to height radiologists. As achieved by previous studies \cite{usman_volumetric_2020,al-shabi_procan_2022,wu_joint_2018} we only consider the nodules annotated by at least 4 radiologists with a diameter greater or equal than 3 mm. The malignancy of each nodule was rated from 1 to 5 by the radiologist. Following \cite{al-shabi_procan_2022,al-shabi_gated-dilated_2019}, the final malignancy labels is obtained by taking the median value of the ratings of all radiologists and the nodules with a median of 3 were excluded as no benign or malignant assignment can be given for them. This selection results in a total of 656 nodules, among which 352 are malignant (54$\%$), and 304 are benign (46$\%$) (cf. Table~\ref{tab1}). As in \cite{usman_volumetric_2020,kubota_segmentation_2011}, a 50 $\%$ consensus criterion is opted to generate the segmentation mask boundaries of the nodules in order to remove the variability among the radiologists.

\subsection{Data analysis models}
\begin{figure}
\includegraphics[width=\textwidth]{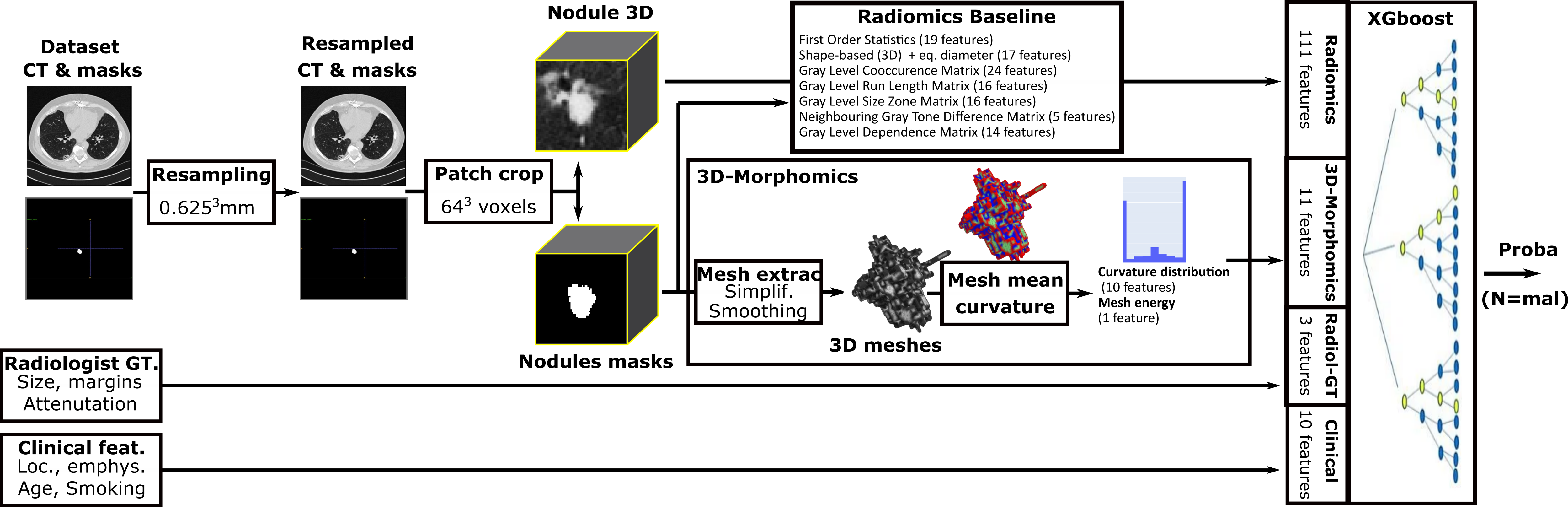}
\caption{The workflow of 3D-morphomics} \label{fig1}
\end{figure}

The 3D-morphomics process reconstructs a 3D mesh from a binary mask of an organ or sub-tissue, and automatically extract curvature features from the mesh. The features are then used for the diagnosis, e.g. the prediction of the malignancy of lung nodules. Thus, the process is composed of 3 main steps (Figure\ref{fig1}).\\

\subsubsection{Pre-processing}
The slice spacing vary according to center and CT acquisition apparatus, for example in the NLST and LIDC collections, ranging from 0.45 to 5 mm. Masks and CT volumes are thus re-sampled to cubical voxel size $0.625^3$ mm using nearest-neighbor method and cubical spline respectively. Patches of $64\times64\times64$  voxels centered on the barycenter of mask are then extracted.\\

\subsubsection{3D-morphomics features} 
A mesh is constructed from the boundaries of a 3D arbitrary mask volume of an organ using the Marching Cube of Lewiner algorithm as it resolves ambiguities and guarantees topologically correct results. The reconstructed mesh is then simplified to improve triangles quality by collapsing short edges, splitting long edges, removing duplicated vertices, duplicated faces, and zero area faces.\\
As the voxels resolution induce noise in the estimation of the curvature, we tested Poisson's, Taubin's, and z axis anisotropic smoothing of the meshes. None of them resulted in performance improvements and they were hence discarded. Both Gauss or mean discrete curvature are extracted \cite{meyer_discrete_2003}. The mean curvature 
$K_i$ at a vertex $i$ is defined via the Steiner approximation \cite{meyer_discrete_2003} \ref{mean_curv}:
\begin{equation}\label{mean_curv}
    K_i= \frac{1}{4}\sum_{ij}\theta_{ij}l_{ij},
\end{equation} 
where $\theta_{ij}$ is the edge dihedral angle and $l_{ij}$ is the edge length as defined above (and the sum is taken over halfedges extending from $i$). The probability distribution (normalized to unit area) of the mean curvature $K_G$ on the mesh are computed and binned to 10 range values distribution histograms (with fixed max and min of -0.2 and 0.2), together with the associated mesh’s energy given by the formula \ref{equation_energy}

\begin{equation}\label{equation_energy}
    E(M)= \int_{M} |K_G| dM
\end{equation} 
The windowing and number of bins of the distribution constitute the main hyper-parameters of the method that can be tuned as classically. It provides the 11 morphological features that are used for the classification. The number of bins shall be chosen as a function of the sample size in order to avoid curse of dimensionality problems.\\

\subsubsection{Radiomics} 
Following \cite{van_griethuysen_computational_2017,aerts_decoding_2014}, 111 radiomic features are extracted, including:
\begin{itemize}
    \item 19 of first order statistics, 
    \item 17 3D shape-based (one additional redundant feature of "volume" given computed by counting the voxels volumes), 
    \item 24 Gray Level Cooccurence Matrix (GLCM), 
    \item 16 Gray Level Run Length Matrix (GLRLM), 
    \item 16 Gray Level Size Zone Matrix (GLSZM), 
    \item 5 Neighbouring Gray Tone Difference Matrix (NGTDM), 
    \item 14 Gray Level Dependence Matrix (GLDM).\\ 
\end{itemize}
    
\subsubsection{Clinical features}
Clinical features consist in patient information provided by NLST database: age ('age'), pack year ('pkyr’), Average number of cigarettes per day ('smokeday’), Average number of cigarettes per day, Cigarette smoking status ('cigsmok', current vs former), Age at smoking onset ('smokeage'), Total years of smoking ('smokeyr'), family antecedent ('family\_ant', if lung cancer was diagnosed in brother (+1), sister(+1), mother(+1), father(+1)), localization of the nodule on the height axis ('localization’, lobe: down, middle, up), Emphysema PSE (‘emph\_PSE’, no, mild, substantial), Emphysema score (‘emph\_sc’, 0-4).\\

\subsubsection{Radiologist GT features} 
Radiologist GT features consist in the longest axis diameter measured by our radiologist using semi-automated tool, and two qualitative features provided by NLST radiologists: the "Margins" (smooth, poorly defined, spiculated), and "Attenuation" (mixed, Ground Glass, soft tissue, fat, water).\\

\subsubsection{Classifier} 
The inference model is a gradient boosted decision tree (XGBoost \cite{chen_xgboost_2016}). The hyper-parameters of XGBoost are tuned on the train-validation set of NLST with a 80-20\% split using Bayesian descent on a binary logistic loss. The space of exploration with their range is: 

\begin{itemize}
    \item 'max\_depth'[3-18],'gamma' [0,9], 
    \item 'reg\_alpha'[1e-5, 1e-4, 5e-4, 1e-3, 5e-3, 1e-2, 5e-2, 0.1, 0.5, 0.7, 1, 2, 3, 5, 10, 20, 50, 80, 100], \item 'reg\_lambda'[0,1],
    \item 'colsample\_bytree'[0.5,1], 
    \item 'min\_child\_weight'[0-10],
    \item 'subsample'[0.5,1] 
\end{itemize}
with a number of estimator set to 180. The model is retrained on the train set with the set of optimal parameters with a learning rate of 0.01 and 1000 estimators. The input imbalance (scale\_pos\_weight) was set to the observed imbalance. The models are then validated against the independent test set of NLST or the LIDC cohort, and models AUC are compared using unpaired 2-sided Welch t-test with p-value p<0.05 on n=5000 bootstrap samples of the test sets.\\

\subsubsection{Dependencies and computation time:} PyRadiomics 3.0.1, Scikit-Learn 0.24.2, Pymesh 0.3, Xgboost 1.5.2, OpenMesh, Scikit-Image 0.18.3,  hyperopt 0.2.7, SimpleITK 2.2.1. The computation of 3D-morphomics is light and takes 1.2s per nodule on average on a standard CPU (IntelCore i7-6700K CPU@4.00GHz).

\section{Results}

\subsection{3D-morphomics}\label{methods}
\begin{figure}
\includegraphics[width=\textwidth]{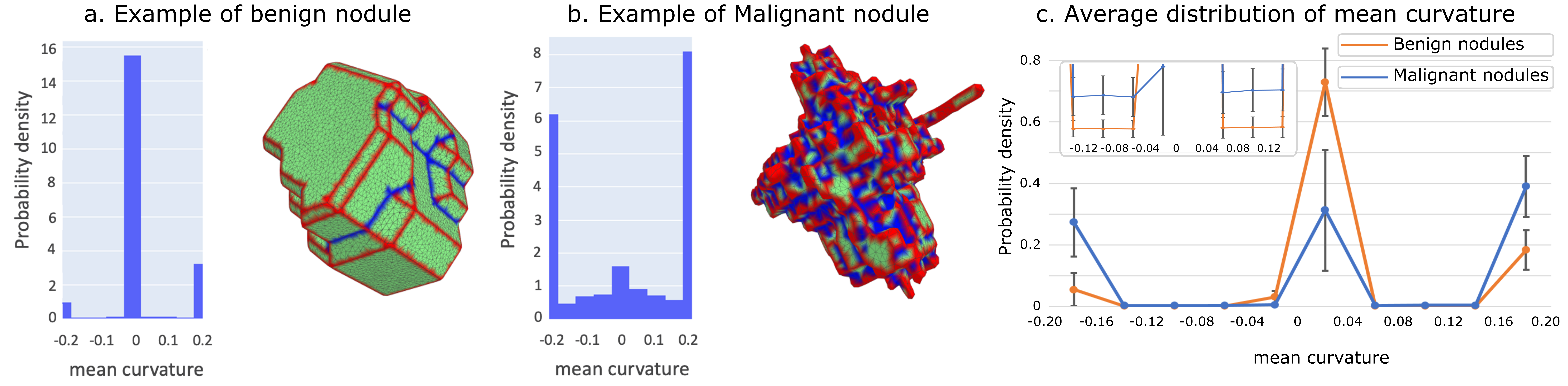}
\caption{Examples of Benign and Malignant nodules curvature distribution and meshes (a., b., the mean curvature values at each vertex are color coded from blue-negative, green-close to zero, red-positive) and (c.) the average distributions of mean curvature across all NLST set of malignant and benign nodules with error bars representing their standard deviations (enlargement for small values is given inset).} \label{fig2}
\end{figure}

A visualisation of examples and basic statistical analysis of the curvature distribution features is illustrated in Figure \ref{fig2} for the whole set of nodules of NLST dataset. As expected, the curvature distributions obtained from malignant and benign nodules are very different, the malignant nodules displaying a clear over-representation of high negative and positive curvature values,  while benign nodules display a clear over-representation of low curvature values. This shows that the computed features effectively captures the shape irregularities such as the spiculations that are discriminative features of malignancy for clinician. \\
Considering each curvature features independently, the mean of the distributions of the malignant and benign are highly significantly different for each feature (for all features we have $p<<0.001$ with a two-sided unequal variance t-test on the mean value). It can also be observed from Figure \ref{fig2}c. that in both cases the distributions have an overall bias toward positive curvatures, as a direct consequence of the fact that nodules are isomorphic to the sphere. 

The study of Gaussian curvatures is not shown here and, as illustrated by Leonardi \cite{leonardi_modelisation_2014} (p.100) combining Gaussian and mean curvatures allows to encode more information about the shape such as valleys, ridges, saddles, pick and peat, hence suggesting some possible improvements on the present features for more subtle classifications.

\subsection{Lung nodule diagnosis performances of 3D-morphomics}
Following the process exposed in \ref{methods}, we trained 5 different models on the NLST train-valid set corresponding to the set of 3D-morphomics, radiomics, clinical, radiologist GT, and the combination of 3D-morphomics with radiomics features. The performances estimated on the NLST test set and on the LIDC dataset are resumed in Table~\ref{tab2}.

\begin{table}
\centering
\caption{Models performances on NLST and LIDC test sets: AUC, sensitivity and specificity at the maximum of Youden index, accuracy, 5000 bootstraps mean AUC $\pm$ standard deviation and 95$\%$ Confidence Interval (bold: best automated models).}\label{tab2}
\begin{tabular}{|l|l|l|l|l|l|}
\hline
Models (test set) &  AUC & Sens. & Spec. & Acc. & AUC Mean$\pm$Stdev[95$\%$CI]\\
\hline
3D-morphomics (NLST) & 0.964 &  90.7 & 91.1 &  0.94 & 0.964$\pm$0.006[0.952 0.976]\\
Radiomics (NLST)   & 0.976 & 92.7 & 93.6 & 0.94 &  0.977$\pm$0.004[0.968 0.984] \\
Clinical feat. (NLST)  & 0.58 &  64.2 & 52.0 & 0.61 &  0.58$\pm$0.023[0.534 0.625]\\
Radiologist GT feat. (NLST) & 0.979 & 93.3 & 93.4 & 0.93 & 0.979$\pm$0.004[0.971 0.986]\\
Brock NLST model \cite{winter_external_2019} (NLST)  & 0.826 &  72.0 &  82.0 & 0.81 &  0.826$\pm$0.02[0.786 0.864]\\
3D-morphomics+Radiomics (NLST)  & \textbf{0.978} & 92.7 & 94.7 & \textbf{0.95} &  \textbf{0.979$\pm$0.004[0.971 0.986]}\\
3D-morphomics (LIDC)   & 0.906 & 84.0 & 85.8 & 0.84 & 0.906$\pm$0.012[0.883 0.928]\\
Radiomics (LIDC)   & 0.956 & 88.0 & 90.7 & 0.89 & 0.956$\pm$0.007[0.941 0.969]\\
3D-morphomics+radiomics (LIDC)  & 0.958 & 91.7 &  87.1 & 0.90 & 0.958$\pm$0.007[0.944 0.971]\\
\hline
\end{tabular}
\end{table}

\begin{figure}
\includegraphics[width=\textwidth]{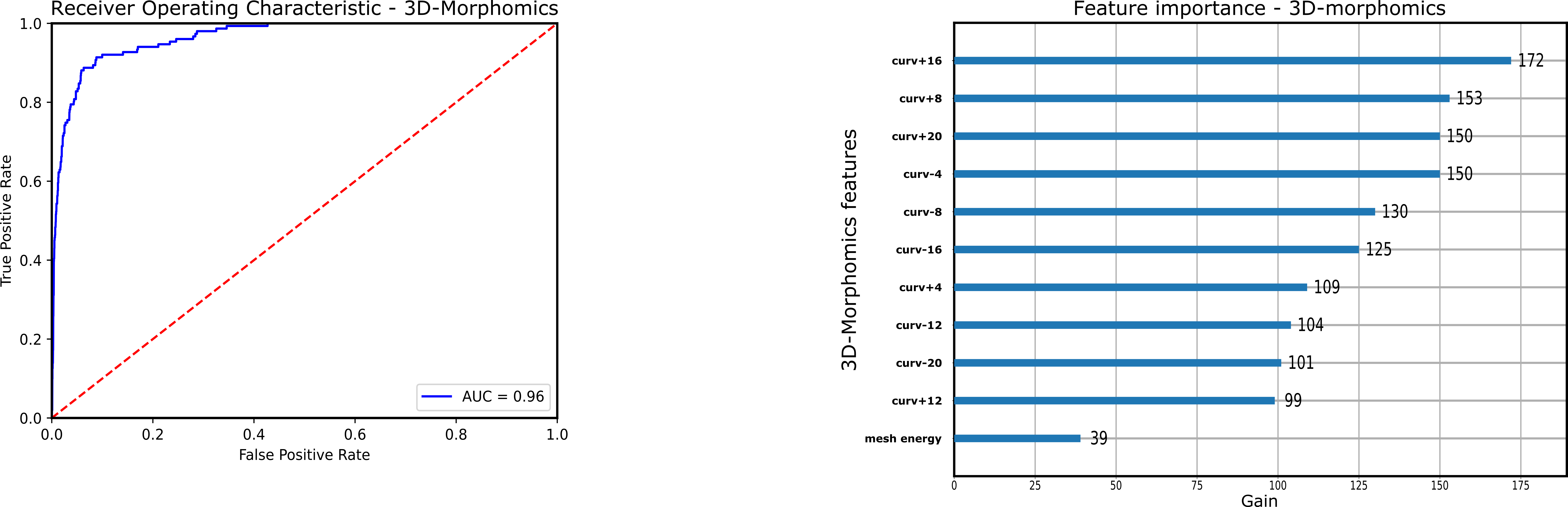}
\caption{\textbf{Performance and feature importance of 3D-morphomics model:} (left) ROC Curves of the 3D-Morphomics model on NLST test set. (right) XGBoost feature importance on NLST test set.} \label{fig4}
\end{figure}

The \textbf{3D-morphomic model} exhibits satisfying performances with an AUC of 0.964, in the sens that it significantly outperform the NLST Brock model \cite{winter_external_2019} (Welch t-test n=5000, p<0.05), and indirectly the AUC of 0.82 obtained by Choi et al. using morphological features on LIDC \cite{choi_reproducible_2021}. \\
The high positive curvature have the highest predictive power. The importance of features for the 3D-morphomics model is given in Figure \ref{fig4}. \\

\begin{figure}
\includegraphics[width=\textwidth]{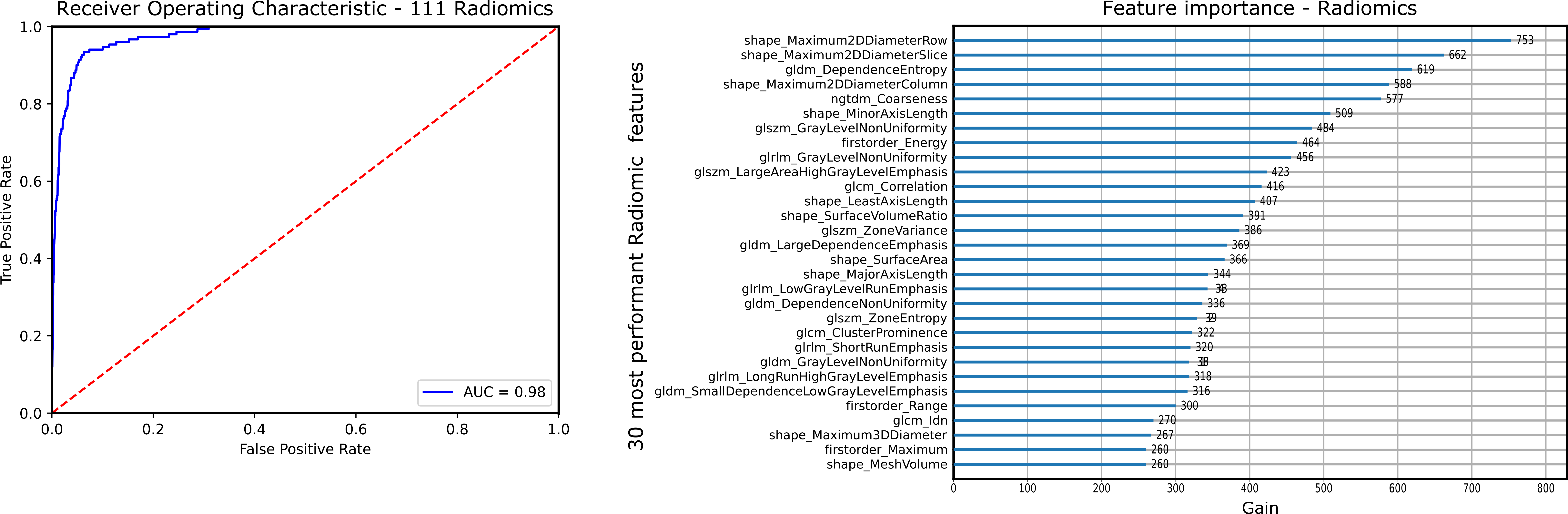}
\caption{\textbf{Performance and feature importance of 111 radiomics model:} (left) ROC Curves of the Radiomics model on NLST test set. (right) XGBoost feature importance (30 highest importance among the 111) on NLST test set.} \label{fig5}
\end{figure}

As expected, the \textbf{radiomics model}, which gathers 111 features representing size, classical 3D shape and luminance patterns functions (first order stat, GLCM, GLRLM, GLSZM, NGTDM, GLDM) provides even higher performances with an AUC of 0.976. Four among the 6 highest predictive features are 2D shape features, while the 2 left are NGTDM and GLDM features. The importance of features for the radiomics model is given in Figure \ref{fig5}.\\

\begin{figure}
\includegraphics[width=\textwidth]{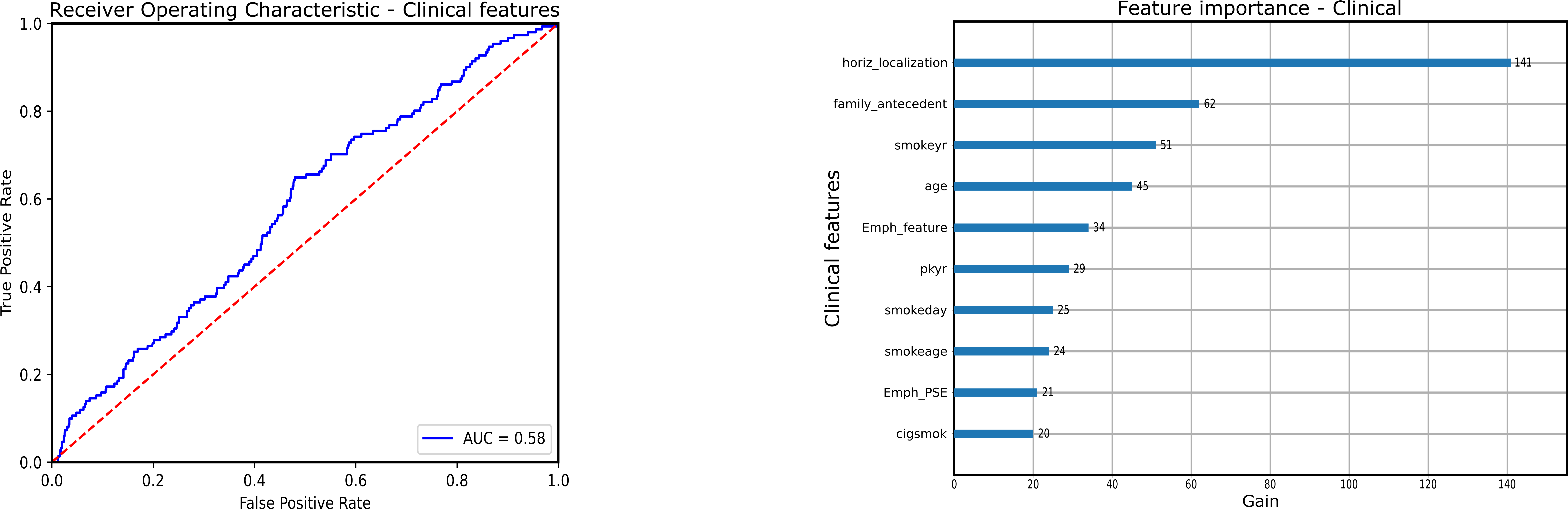}
\caption{\textbf{Performance and feature importance of clinical model:} (left) ROC Curves of the Clinical model on NLST test set. (right) XGBoost feature importance on NLST test set.} \label{fig6}
\end{figure}

The \textbf{clinical model} with features partially overlapping with the PANCAN model here applied at the nodule level, provides poor performance, indicating that clinical informations such as age, smoking habits, family antecedent, emphysema, have low impact on the prediction of a lesion malignancy. The nodule horizontal axis localization has far the most important predictive power, followed by family antecedent, the number of smoking years, and then age. The importance of features for the clinical model is given in Figure \ref{fig6}. \\ 

\begin{figure}
\includegraphics[width=\textwidth]{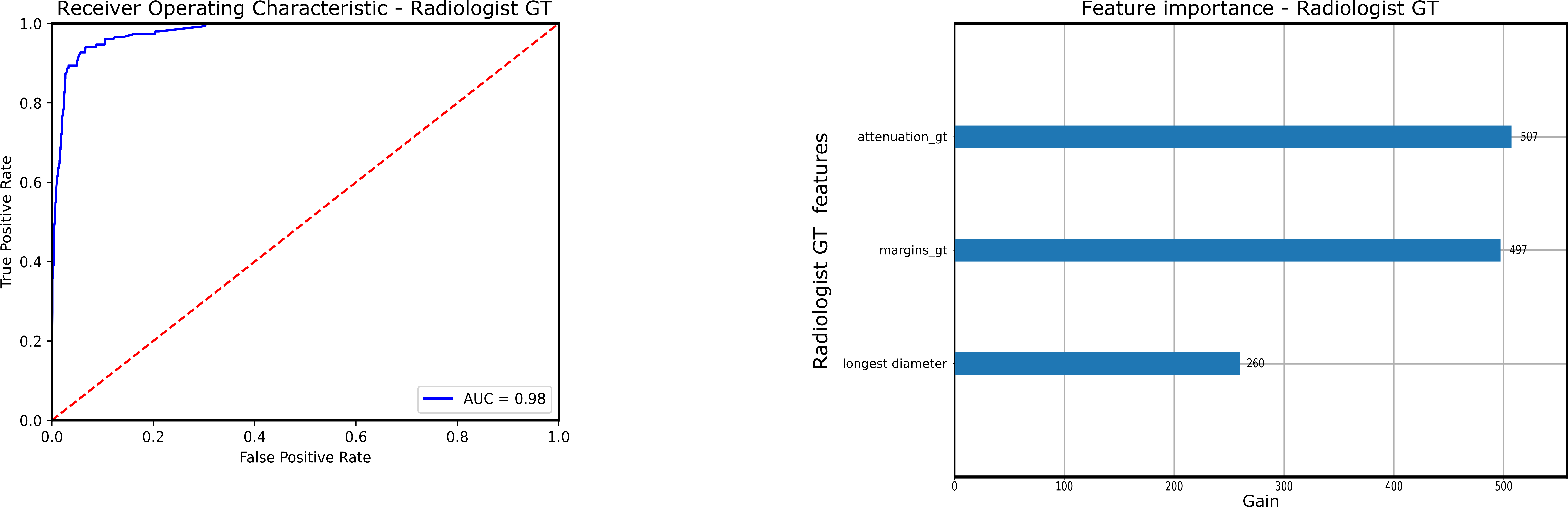}
\caption{\textbf{Performance and feature importance of Radiologist GT model:} (left) ROC Curves of the Radiologist GT model on NLST test set. (right) XGBoost feature importance on NLST test set.} \label{fig7}
\end{figure}

As expected also, the \textbf{radiologist GT model}, based on 3 main diagnostic criterions of size,  qualitative margins (shape) and attenuation (texture) of a nodule, gives a high AUC of 0.979. As a surprise, attenuation and margins had about twice greater predictive power than the size (given by the longest diameter). The importance of features for the radiologist GT model is given in Figure \ref{fig7}. \\

\begin{figure}
\includegraphics[width=\textwidth]{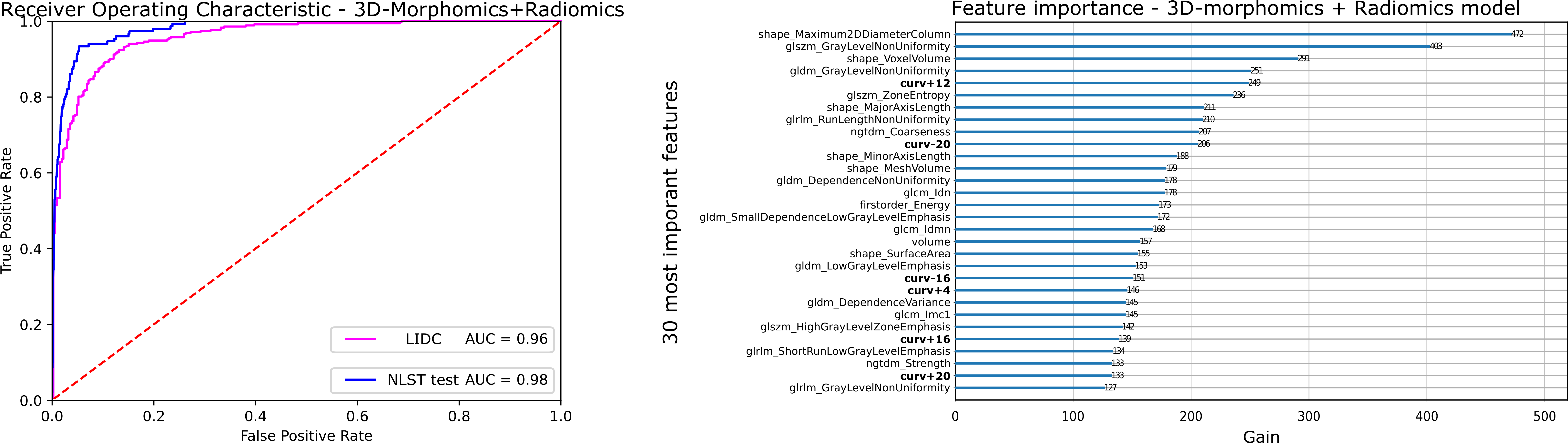}
\caption{Main results: (left) ROC Curves of the 3D-Morphomics+Radiomics model on NLST test set and LIDC dataset. (right) XGBoost feature importance (30 highest) on NLST test set. Curvature distribution features are in bold. } \label{fig3}
\end{figure}

The main result of the paper is that the \textbf{combination of radiomics and 3D-morphomics} gives a performance of 0.978, as illustrated in Figure \ref{fig3}. As shown in the graphic of feature importance, 6 curvature distribution features resides among the 30 features with highest predictive power. Within the top 5 of features with highest predictive power we can observe 2 shape feature (2D size and volume) 2 texture feature (grey level non uniformity of GLSZM and GLDM) and a 3D-morphomics feature of high positive curvature. We also tested the Brock NLST model \cite{winter_external_2019}, which combines some clinical and annotators GT features into a logistic model, and obtained an AUC of 0.826, lower than the originally reported but partly confirming recent results \cite{chetan_developing_2022}.  \\

In order to validate the model and results on a publicly available independent cohort, the 3D-morphomics and 3D-morphomics+radiomics model are tested on the LIDC dataset, giving an AUC of 0.906 which is higher than the 0.82 obtained with other morphological feature by Choi et al. \cite{choi_reproducible_2021} and of 0.958 respectively (cf. Table~\ref{tab2}). The latter score is ranking second at the associated LIDC characterization challenge, just after \cite{al-shabi_procan_2022}, and despite the clear disadvantage that our model was not trained on LIDC as opposed to the other solutions. The result of unpaired 2-sided Welch t-test (n=5000, p<0.05) rejects the null hypothesis that 3D-morpho+Radiomics and Radiomics have equal mean on both NLST and LIDC test set.\\

We conclude that the good performance of the 3D-morphomics+radiomics model generalize satisfyingly to other data, and is robust to important dataset variations: while NLST is a low dose CT dataset, LIDC is not.   
We remark that the malignancy GT of LIDC is much more uncertain with respect to the true malignancy status of a nodule than the accurate estimation given by the biopsy results for malignancy and by the up to 7 year follow up for benignity. Notably, the malignancy GT of LIDC contains all the FP and FN of radiologists, e.g. the False Discovery Rate of radiologists on our NLST dataset is 5.9$\%$. Hence, the results obtained on the test set of NLST and trained on NLST against the biopsy derived GT, gives a more accurate estimation on the real performances of our model than the results obtained with the LIDC set that are biased by radiologist diagnostic errors.   

\section{Conclusions}
It is a surprise that morphomics and radiomics approaches, relying on pre-designed features, can provide equivalent or even out-perform deep network models on a medical diagnosis task \cite{al-shabi_lung_2019,AlShabi2019,2022}. Given previously reported high performances of deep models on the NLST cohort \cite{baudot_development_2022}, those results can only leave the hope that a combination of deep and 3D-morphomics solutions will achieve even better performances. \\
This work establishes the distribution curvature as efficient features for predicting lung nodule malignancy and that the nodules shape have a predominant predictive power beyond the classical size used in everyday clinical practice at the image of the Lung-RADS score. Moreover, the method presented here based on a new family of observable can be applied directly to arbitrary computer aided diagnosis task, and is currently applied to liver fibrosis stage diagnostic where the nodularity of the liver surface is symptomatic. The model is quite generic in order to detect morphological deformations due to pathologies, but suffers from reducing the morphology to a 1D vector of curvature distribution, which erase some 2D spatial information. To overcome this limitation, an appealing perspective could be to use Graphical Neural Network classifiers taking as input the 2D network of the meshes, as explored in \cite{qiu_dense_2021}. 

\section*{Acknowledgments}
This work was partly supported by ANR plan de Relance. The authors warmly thank Sebastien poulot, Yael Freguier, Grégoire Sergeant Perthuis, Vincent Bobin, Mahaut Macrez, Afef Baili-Laya, Jean-Luc Mari and all the iBiopsy team for discussions, help and supports.

\bibliographystyle{unsrt}  
\bibliography{bib_morphomics}

\end{document}